# Pronóstico del índice planetario Kp usando modelos autoregresivos

Arian Ojeda González [1,2], Clezio Marcos Denardini [1], Siomel Savio Odriozola [1,2], Reinaldo Roberto Rosa [1] y Odim Mendes Jr. [1]

[1] *Instituto Nacional de Pesquisas Espaciais-INPE, Brasil. E-mail: ojeda.gonzalez.a@gmail.com , clezio.denardin@inpe.br, reinaldo@lac.inpe.br  odim.mendes@inpe.br*
[2] *Instituto de Geofísica y Astronomía, La Lisa, La Habana, Cuba E-mail: siomel@gmail.com*

**Resumen**

El índice geomagnético Kp se deriva del índice K a partir de las mediciones de trece estaciones localizadas alrededor de la Tierra entre las latitudes geomagnéticas 48◦ y 63◦. Este índice se procesa cada tres horas, es cuasi-logarítmico y estima la actividad geomagnética. Los valores de Kp están dentro de un rango de 0 a 9 y conforman un conjunto de 28 valores discretos, se utiliza como uno de los parámetros de entrada en muchos modelos ionosféricos y magnetosféricos. El objetivo de este trabajo es utilizar datos históricos del índice Kp para desarrollar una metodología que permita hacer un pronóstico del mismo en un intervalo de tiempo, como mínimo, de tres horas. Se prueban cinco diferentes modelos de pronóstico de los índices geomagnéticos Kp y ap. Se utilizan como datos de entrada a los modelos, una serie temporal de valores del índice Kp desde 1932 hasta el 15/12/2012 a las 21:00 horas UT. La finalidad del modelo es pronosticar los tres valores posteriores al último valor medido del índice Kp (las próximas 9 horas). El modelo AR resulta ser el de menor costo computacional y ofrece buenos resultados. El modelo ARIMA es eficiente para la predicción del índice Kp en condiciones de perturbación geomagnética. Este trabajo ofrece una forma rápida y eficiente de hacer una predicción del índice Kp, sin necesidad de usar datos de satélites que muchas veces demoran en ser publicados. Aunque se informa que los resultados del pronóstico son mejores cuando se utilizan datos de satélites. Según datos publicados, la correlación lineal entre los valores pronosticados y los valores reales está entorno de un 77 %, valor que es mejor que el 68.5% obtenido en este trabajo. Sin embargo, teniendo en cuenta que se trabajó solamente sobre la serie temporal estocástica del Kp, este valor de correlación puede considerarse satisfactorio.

**Palabras clave:** Clima espacial, índice Kp, modelos autoregresivos, pronóstico de series temporales

## Planetary Kp index forecast using autoregressive models

**Abstract**

The geomagnetic Kp index is derived from the K index measurements obtained from thirteen stations located around the Earth geomagnetic latitudes between 48◦ and 63◦. This index is processed every three hours, is quasi-logarithmic and estimates the geomagnetic activity. The Kp values fall within a range of 0 to 9 and are organized as a set of 28 discrete values. The data set is important because it is used as one of the many input parameters of magnetospheric and ionospheric models. The objective of this work is to use historical data from the Kp index to develop a methodology to make a prediction in a time interval of at least three hours. Five different models to forecast geomagnetic indices Kp and ap are tested. Time series of values of Kp index from 1932 to 15/12/2012 at 21:00 UT are used as input to the models. The purpose of the model is to predict the three measured values after the last measured value of the Kp index (it means the next 9 hours values). The AR model provides the lowest computational cost with satisfactory results. The ARIMA model is efficient for predicting Kp index during geomagnetic disturbance conditions. This paper provides a quick and efficient way to make a prediction of Kp index, without using satellite data. Although it is reported that the forecast results are better when satellite data are used. In the literature we find that the linear correlation between predicted values and actual values is 77 %, which is better than the 68.5% obtained in this work. However, taking into account that our results are based only on Kp stochastic time series, the correlation value can be considered satisfactory.

**Keywords:** Space weather, Kp index, autoregressive models, time series forecasting.

*Arian Ojeda González, Instituto Nacional de Pesquisas Espaciais -INPE, Brasil, E mail: ojeda.gonzalez.a@gmail.com*

**1. Introducción**

El nivel de perturbación global del campo geomagnético producto de los disturbios interplanetarios (ej. eyección de masa coronal o CME, por sus siglas en inglés: *Coronal Mass Ejection*) que llegan a la Tierra desde la atmosfera solar es registrado a través del índice Kp. Este índice tiene una frecuencia de tres horas, o sea en un día de observación se reportan ocho valores de Kp. Este índice geomagnético se deriva del índice K obtenido de las mediciones de trece estaciones localizadas alrededor de la Tierra entre las latitudes geomagnéticas 48° y 63°. Los valores de Kp están comprendidos dentro de un rango de 0 a 9 y conforman un conjunto de 28 valores discretos: 0o, 0+, 1-, 1o, 1+,..., 8-, 8o, 8+, 9-, 9o. El valor mínimo (0) representa condiciones magnéticas extremadamente calmas y el valor máximo (9o) una perturbación geomagnética muy alta. Los valores finales de Kp son publicados con un atraso de aproximadamente dos meses, lo que representa una dificultad para estudios a tiempo real del clima espacial. En este sentido se han desarrollado algoritmos para calcular casi en tiempo real un estimado de Kp (ej. Gehred et al., 1995; Takahashi et al., 2001).

La modelación de una serie de tiempo es un problema estadístico, donde las observaciones varían de acuerdo con alguna distribución de probabilidad sobre una función subyacente del tiempo. Los modelos más simples son el de Nivelado Exponencial y el de la Media Móvil, así como aquellos derivados de tecnologías de Redes Neuronales, de Algoritmos Genéticos y otros enfoques que utilizan combinación de estas técnicas en la forma de sistemas híbridos (Shumway y Stoffer, 2006).

La previsión de Kp en hasta 3 horas a partir de datos del viento solar utilizando Redes Neuronales fue desarrollado por Boberg et al. (2000); Wing et al. (2005). Siguiendo estas ideas, en este trabajo se desarrolla un algoritmo de predicción del índice Kp con la finalidad de poder incorporarlo a un sistema de monitoreo y pronóstico magneto-ionosférico. En este caso, a diferencia del método de Boberg et al. (2000) será realizado un análisis a partir de la serie temporal del Kp, útil en aquellos casos donde no se tenga un control administrativo sobre los datos satelitales del viento solar; ganándose así, autonomía operativa.

El índice Kp es importante porque se utiliza como uno de los parámetros de entrada en muchos modelos ionosféricos y magnetosféricos que permiten tomar medidas preventivas para evitar o reducir daños económicos en las redes de comunicación, energía y satelitales.

El objetivo de este trabajo es utilizar datos históricos del índice Kp para desarrollar una metodología que permita hacer un pronóstico del mismo en un intervalo de tiempo, como mínimo, de tres horas.

El trabajo se ha dividido en tres secciones. En la Sección 2 presentaremos la forma en que se organizó la base de datos del Kp y los métodos utilizados. En la Sección 3 se presentan los resultados y realizamos una discusión de los mismos. Por último, en la Sección 4 colocamos las conclusiones del trabajo.

**2. Materiales y métodos**

Para realizar una predicción de un parámetro físico se necesita una base de datos lo suficientemente extensa y coherente del parámetro en cuestión. Generalmente, se procede a organizar la base de datos eliminando informaciones innecesarias. Muchas veces existen *string* e inexistencias de valores (*gap*) que debemos eliminar o llenar con números reales. En este trabajo hicimos subrutinas de programación para organizar las bases de datos, porque el índice Kp tiene la dificultad de ser publicado en forma de un *string*. Y para trabajar con él, debemos transformarlo en un valor representativo del conjunto de los números reales.

Posteriormente es necesario implementar métodos numéricos que nos permita procesar la base de datos y obtener los resultados propuestos. En este trabajo utilizaremos cinco métodos para hacer el pronóstico del índice Kp. Los resultados derivados de los cinco métodos son contrastados posteriormente.

**2.1 Materiales**

El Kp es un índice trihorario (que se procesa cada tres horas) cuasi-logarítmico que estima la actividad geomagnética. Durante un día se calculan ocho valores de Kp en intervalos de tres horas en tiempo universal (*universal time, UT*): 00:00, 03:00, 06:00, 09:00, 12:00, 15:00. 18:00, 21:00 UT. Estos valores se obtienen del centro mundial de datos geomagnéticos, de la universidad de Kyoto (1932-2012) (http://wdc.kugi.kyotou.ac.jp/kp/index.html).

El formato de estos datos se representa en la Tabla I. La primera columna informa la fecha (año, mes y día); en la segunda y tercera columnas se representan los ocho valores del Kp y el valor total resultante de la suma de ellos. En las restantes columnas se muestran los valores del índice ap y Ap respectivamente. La forma en que se derivan los valores del índice ap a partir del Kp se muestra en la Tabla II. En cada columna se muestra el valor del Kp y su respectivo valor del índice ap, estos corresponden a una escala logarítmica.

Los datos de la Tabla I son almacenados como un único archivo en formato ASCII. Para seleccionar los datos fue creada una rutina de programación que permite seleccionar los dados del Kp ya almacenados y transformarlos en números reales, esto es, en valores reales únicos representativos. Lo ideal es probar los métodos de pronóstico utilizando los datos del índice ap y posteriormente, para visualizar los resultados, retomar los valores del índice Kp según la propia Tabla II. Sin embargo, deben hacerse algunas alteraciones por causa que la escala logarítmica del ap introduce algunos errores en el pronóstico.

**Tabla I. Formato de la base de datos de los índices Kp, ap y Ap obtenidos del centro mundial de datos geomagnéticos de la Universidad de Kyoto (1932-2012). Los dados del Kp utilizados en este trabajo son los ocho valores diarios calculados en intervalos de tres horas (00:00, 03:00, 06:00, 09:00, ... , 21:00)**

| YYYYMMDD | Kp[8] | Sum | ap[8] | Ap |
|---|---|---|---|---|
| 19320101 | 3+ 3- 2+ 3- 3- 3- 3- 3+ | 24- | 18 12 9 12 18 12 18 18 | 15 |
| 19320102 | 4- 4- 3+ 4- 3+ 5- 3 5 | 30+ | 22 22 18 22 18 39 15 48 | 26 |
| 19320103 | 3+3+ 3  1   2+ 2- 3- 2 | 19 | 18 18 15 4 9 6 12 7 | 11 |
| ............ | .............................. | | ................................ | ... |
| ............ | .............................. | | ................................ | ... |
| ............ | .............................. | | ................................ | ... |
| 20121213 | 2- 1 0 0  0+ 0+ 1- 0+ | 4+ | 6 4 0 0 2 2 3 2 | 2 |
| 20121214 | 1 1 1- 0+ 0  0+ 1  2- | 6 | 4 4 3 2 0 2 4 6 | 3 |
| 20121215 | 2 3- 2- 2- 3  2  2+ 2+ | 18- | 7 12 6 6 15 7 9 9 | 9 |

**Tabla II. Índice ap a partir del índice Kp.**

| Kp: | 0o | 0+ | 1- | 1o | 1+ | 2- | 2o | 2+ | 3- | 3o | 3+ | 4- | 4o | 4+ |
|---|---|---|---|---|---|---|---|---|---|---|---|---|---|---|
| ap: | 0 | 2 | 3 | 4 | 5 | 6 | 7 | 9 | 12 | 15 | 18 | 22 | 27 | 32 |
| Kp: | 5- | 5o | 5+ | 6- | 6o | 6+ | 7- | 7o | 7+ | 8- | 8o | 8+ | 9- | 9o |
| ap: | 39 | 48 | 56 | 67 | 80 | 94 | 111 | 132 | 154 | 179 | 207 | 236 | 300 | 400 |

**Tabla III. Escala utilizada para linealizar el índice Kp.**

| Kp: | 0o | 0+ | 1- | 1o | 1+ | 2- | 2o | 2+ | 3- | 3o | 3+ | 4- | 4o | 4+ |
|---|---|---|---|---|---|---|---|---|---|---|---|---|---|---|
| Kp-lineal: | 0 | 0.25 | 0.75 | 1. | 1.25 | 1.75 | 2. | 2.25 | 2.75 | 3. | 3.25 | 3.75 | 4. | 4.25 |
| Kp: | 5- | 5o | 5+ | 6- | 6o | 6+ | 7- | 7o | 7+ | 8- | 8o | 8+ | 9- | 9o |
| Kp-lineal: | 4.75 | 5. | 5.25 | 5.75 | 6. | 6.25 | 6.75 | 7. | 7.25 | 7.75 | 8. | 8.25 | 8.75 | 9. |

**2.2. Métodos**

Como la escala del ap es logarítmica y los métodos de pronóstico utilizarán polinomios de interpolación, el tiempo de cálculo computacional empleado sería grande si se implementase los diferentes métodos de pronósticos directamente sobre el índice ap. Por esta razón, el Kp fue transformado a una escala lineal de números reales, según el criterio mostrado en la Tabla III, para realizar el pronóstico. Un método donde también se linealiza el Kp fue utilizado en el trabajo de Wing et al. (2005, p. 3). La Tabla III es similar a la Tabla II, la única diferencia radica en que el valor del índice ap ha sido substituido por un Kp-lineal con valores predefinidos.

Para convertir los valores reportados de Kp como se muestra en la Tabla III, se creó una sub-rutina de programación que importa los datos originales en formato ASCII como variables de cadena de caracteres (*string*) y los transforma a números reales siguiendo el criterio definido en la Tabla III. De esta forma, la serie histórica de los valores del índice Kp queda linealizada desde 1932 hasta el 15/12/2012 a las 21:00 horas UT. Un ejemplo de un intervalo de valores del Kp-linealizado, correspondiente al 2012, se representa en la Figura 1. Es decir, se grafica el Kp-linealizado versus el tiempo para el mes de marzo del año 2012.

Los métodos probados para realizar el pronóstico fueron:
1) Método autoregresivo, conocido en la literatura como AR.
2) Método MA *(moving-average process)*.
3) Combinación de los métodos anteriores, conocido en la literatura como ARMA *(autoregressive moving-*

*average process with AR).*
4) Método ARIMA *(autoregressive integrated moving-average process).*
5) Método conocido en la literatura como FARIMA *(fractional, autoregressive, integrated moving-average process).*

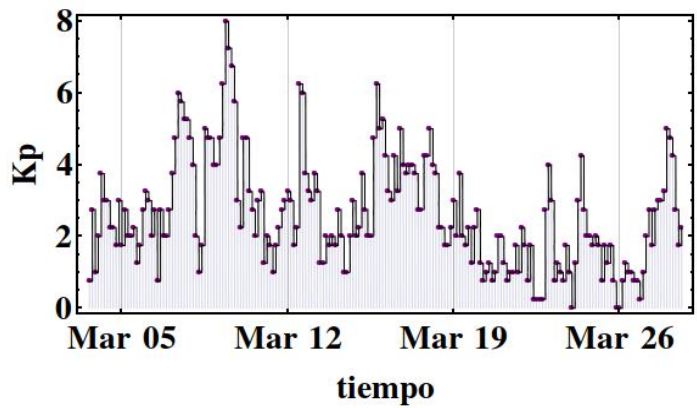

**Fig. 1. Ejemplo de la variación temporal de valores lineales de Kp para una ventana temporal del mes de marzo del año 2012**

Esencialmente, estos métodos se utilizan para modelar la evolución temporal de algún sistema físico dinámico (o en nuestro caso un parámetro que representa a un sistema como es el índice Kp) que representan procesos estocásticos. El sistema físico es estocástico cuando aparece cierto grado de incertidumbre, aunque las condiciones iniciales sean las mismas, el sistema no sigue exactamente la misma sucesión de estados físicos. Dado un estado físico x al tiempo to, no sabemos cuál será el estado físico del sistema al tiempo to+ t (Shumway y Stoffer, 2006).

Todos los métodos utilizados son funciones ya implementadas en el software *Mathematica* en su versión más actual (9.0.0). El formalismo matemático de estos métodos puede encontrarse online como parte de la documentación del software. Estas funciones tienen la sintaxis siguiente:

1- ARProcess[{$a_1,a_2,...,a_p$},v].
2- MAProcess[{$b_1,b_2,...,b_p$},v].
3- ARMAProcess[{$a_1,a_2,...,a_p$},{$b_1,b_2,...,b_p$},v].
4- ARIMAProcess[{$a_1,a2,...,a_p$},d,{$b1,b_2,...,b_p$},v].
5- FARIMAProcess[{$a_1,a2,...,a_p$},d,{$b1,b_2,...,b_p$},v].

Los coeficiente *a* y *b* son parámetros que representan a los métodos AR y MA respectivamente. El parámetro v es la varianza de una serie de tiempo tipo ruido blanco normal. La *d*-ésima diferencia luego de la cual un proceso tipo ARIMA o FARIMA se vuelve un proceso tipo ARMA es especificada a la función mediante el parámetro *d*.

Para hacer un pronóstico de valores no existentes en la serie temporal se hizo uso de la función *TimeSeriesForecast* que aparece recientemente incorporada a la última versión del software *Mathematica*. La sintaxis de esta función es: *TimeSeriesForecast[tproc, data, kspec]*. En este trabajo, la variable de entrada *tproc* define el modelo a usar, o sea los cinco modelos anteriormente mencionados; la variable de entrada *data*, representa la serie temporal histórica del índice Kp ya linealizada y *kspec* el número de valores que quiere pronosticarse. Por ejemplo si *kspec =3* estaremos pronosticando los valores de Kp por las próximas 9 horas.

Con el uso de otra función: *AdjustTimeSeriesForecast[eproc2, forescast2, newdata]* es posible incorporar nuevos datos *(newdata)* al pronóstico previamente hecho usando *TimeSeriesForecas (forescast2)* a medida que avanza el tiempo y el proceso es modelado por uno de los cinco procesos usados en este trabajo *(eproc2)*.

Para justificar físicamente que es viable el uso de modelos autoregresivos para pronosticar el índice Kp es recomendable leer el trabajo de McPherron (1999). Ellos construyen un modelo de pronóstico para el índice planetario diario que mide la fuerza de la perturbación en el campo magnético terrestre, el índice Ap. Este índice se deriva del índice Kp. El modelo de McPherron (1999) también utilizan métodos autoregresivos. Los métodos que usan el comportamiento pasado de una serie de tiempo para predecir los valores futuros se llaman autoregresivos. En principio, de los análisis de persistencia (o memoria de la serie temporal) realizados sobre el Ap, se espera que el valor de mañana sea igual al valor de hoy. Es decir, ellos encuentran autocorrelaciones entre los valores del Ap. Debido a la memoria que existe en la serie temporal del Ap es que McPherron (1999) utilizan con éxito los métodos autoregresivos. Por eso, se tiene la hipótesis que estos métodos deben funcionar para realizar un pronóstico del Kp.

*Ojeda et al.*

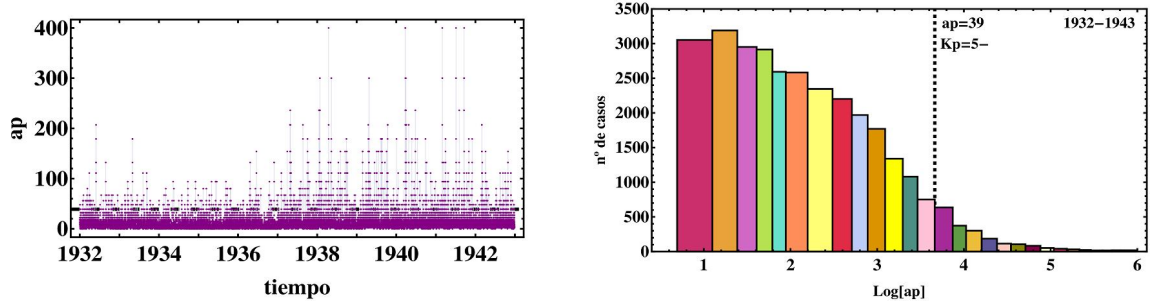

**Fig. 2.** El panel izquierdo muestra el índice ap en el período de 11 años, 1932 - 1943, la línea horizontal discontinua corresponde a un valor ap = 39 o Kp = 5-, a partir del cual el campo geomagnético se considera perturbado. El panel derecho muestra un histograma acumulativo de los valores de ap de este mismo período donde cada barra corresponde a uno de los 28 valores de ap de la Tabla II

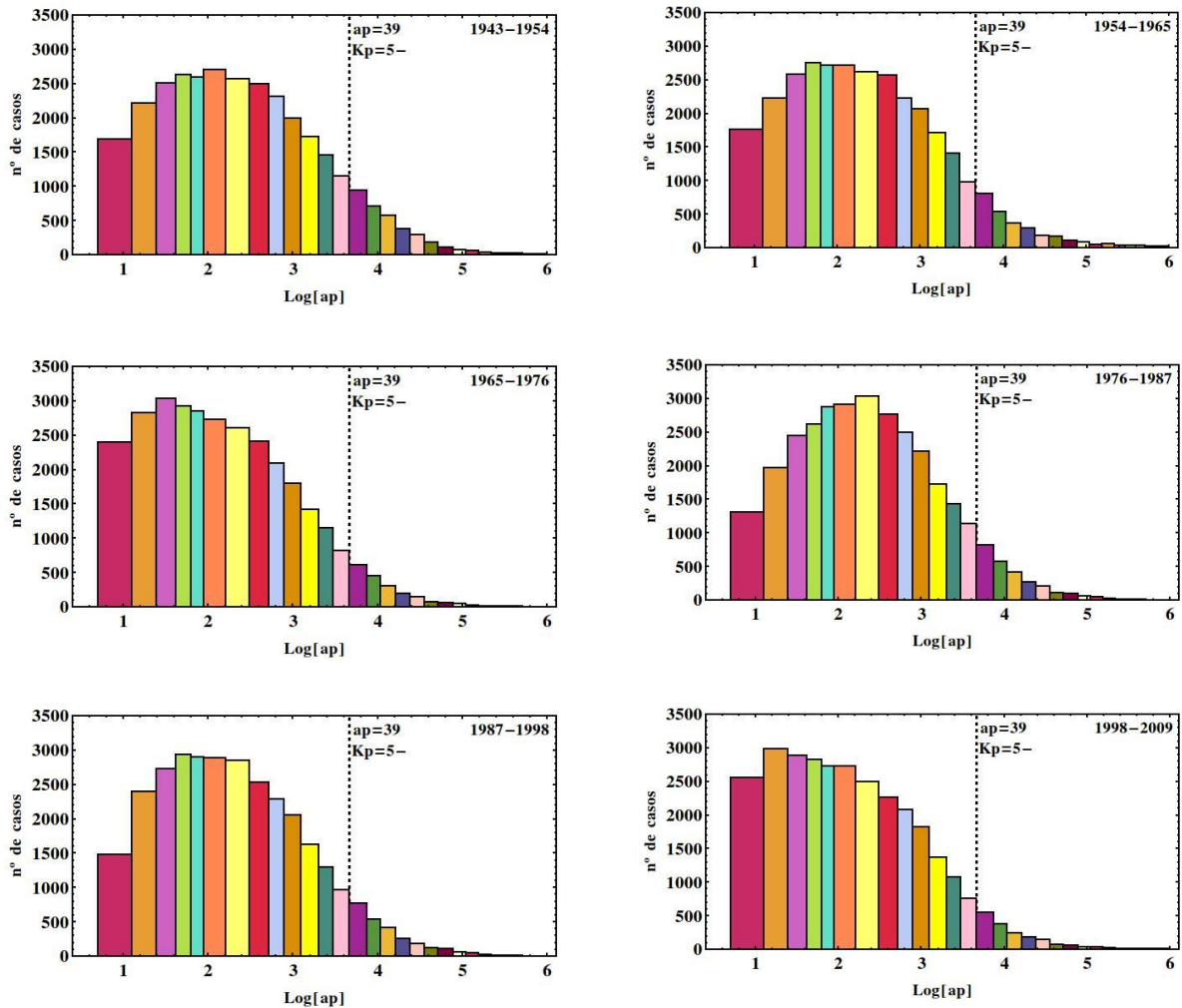

**Fig. 3.** Histogramas acumulativos de la base de datos de los índices ap creada para desarrollar un pronóstico de Kp. No son mostrados los datos desde 2010 hasta el 15/12/2012

**3. Resultados y discusión**

Como los índices Kp y ap son publicados simultáneamente, se puede trabajar directamente con el ap para visualizar los datos. En la Figura 2, panel de la izquierda, se representan los valores del ap en el período de tiempo desde 1932 a las 03:00 horas UT hasta 1943 a las 00:00 horas UT. En el panel de la derecha se muestra el histograma de los valores del Kp en ese intervalo de tiempo (1932-1943). A la derecha de la línea vertical discontinua representada en el histograma, con valores de Kp = 5- y ap = 39, se ubican los valores donde el campo geomagnético estuvo extremadamente perturbado, si seguimos el criterio publicado por la NOAA. Como resultado de este trabajo se obtienen gráficos similares para los períodos, 1943-1954, 1954-1965, 1965-1976, 1976-1987, 1987-1998, 1998-2009, todos ellos se muestran en los seis paneles de la Figura 3. Estas figuras no fueron construidas con la idea de analizar los resultados, el objetivo era explorar y organizar la base de datos del Kp.

Posteriormente, se necesitan validar los cinco métodos. Esto se puede realizar comparando los valores derivados del pronóstico con los datos reales. Este proceso de pronóstico y actualización del mismo es mostrado en el siguiente ejemplo. Inicialmente se toman once años de datos, desde 1987 hasta Enero de 1998. Esos datos son utilizados para hacer un pronóstico de tres horas, o sea del próximo valor de Kp, y a partir del mismo se actualiza la serie de tiempo con los valores reales del Kp para el año 1998. De esta forma es posible construir un gráfico de los valores reales y los obtenidos con el modelo de pronóstico y así evaluar el grado de precisión del modelo empleado.

En la Figura 4 se muestran cinco paneles con los resultados obtenidos utilizando los cinco modelos de pronóstico. Se observa que la correlación es de 68,5 %. Este valor muestra que la serie de tiempo modelada sigue la tendencia de la serie real. Sin embargo, los modelos que utilizan datos de viento solar (Wing et al., 2005) obtienen una correlación de aproximadamente 77 %. No obstante, es un valor que está dentro del intervalo de error esperado y concluimos que los modelos autoregresivos son válidos para efectuar el pronóstico.

En la Figura 4, los pronósticos hechos con los modelos MA y FARIMA no muestran buenos resultados. Esto se destaca en las figuras por una región sombreada en amarillo que representa el área entre la serie de tiempo pronosticada (en rojo) y la serie de tiempo real (en azul) y que resulta ser mayor en los pronósticos que usan los modelos MA (que subestima los valores de Kp en casi toda la ventana de tiempo seleccionada) y FARIMA (que sobrestima los valores de Kp en casi toda la ventana de tiempo seleccionada).

Con los restantes tres modelos AR, ARMA y ARIMA se obtienen buenos pronósticos del índice Kp. El modelo ARIMA tiene un ajuste lineal de la tendencia de la serie temporal lo que funciona mejor para los casos de índices Kp en períodos geomagnéticamente perturbados y por eso consideramos que es el mejor modelo para ser utilizado en estos casos. Los resultados con AR también resultan ser buenos con el agregado de que con este modelo los tiempos de cálculo son mucho menores; lo que consideramos lo coloca como la opción más eficiente a ser utilizada para los fines de pronóstico.

La serie temporal del índice Kp desde 1932 hasta el día 15/12/2012 a las 21:00 horas UT se utilizó como datos de entrada al modelo con el objetivo de hacer un pronóstico del índice en la próximas 9 horas, es decir, de los próximos tres valores de Kp. Esos valores corresponden a 00:00, 03:00 y 06:00 horas del día 16 de diciembre de 2012. En el modelo de predicción también es posible estimar un error cuadrático medio. Este error se puede interpretar como una franja de validez donde debe estar representado el próximo valor numérico real del índice Kp (en el futuro cuando sea reportado).

Aún tenemos un problema que resolver, pues el valor de Kp pronosticado corresponde a un valor real que debe estar entre 0 y 9, pero incluso pudiera dar un valor negativo próximo a cero o mayor que 9. Debemos tener en cuenta estos casos a la hora de realizar el pronóstico. Si el número es menor que cero debemos cambiarlo por 0, y si es mayor que 9 debemos convertirlo en 9. Para los restantes valores entre 0 y 9 debemos tener en cuenta la escala mostrada en la Tabla IV.

**Tabla IV. Escala utilizada para retomar el Kp original después de realizar la predicción.**

| Kp<0.25 = 0o | 0.25≤Kp<0.75 = 0+ | 0.75≤Kp<1 = 1- | 1≤Kp<1.25 = 1o |
|---|---|---|---|
| 1.25≤Kp<1.75 = 1+ | 1.75≤Kp<2 = 2- | 2≤ Kp <2.25 = 2o | 2.25≤ Kp <2.75 = 2+ |
| 2.75 ≤Kp <3 = 3- | 3≤ Kp <3.25 = 3o | 3.25≤ Kp <3.75 = 3+ | 3.75≤Kp <4 = 4- |
| 4≤Kp <4.25 = 4o | 4.25≤Kp <4.75 = 4+ | 4.7≤Kp <5 = 5- | 5≤Kp <5.25 = 5o |
| 5.25≤Kp <5.75 = 5+ | 5.7≤ Kp <6 = 6- | 6≤Kp <5.25 = 6o | 6.25≤Kp <6.75 = 6+ |
| 6.75≤Kp <7 = 7- | 7≤Kp <7.25 = 7o | 7.25≤Kp <7.75 = 7+ | 7.75≤Kp <8 = 8- |
| 8≤Kp <8.25 = 8o | 8.25≤Kp <8.75 = 8+ | 8.75≤Kp <9 = 9- | 9≤Kp = 9o. |

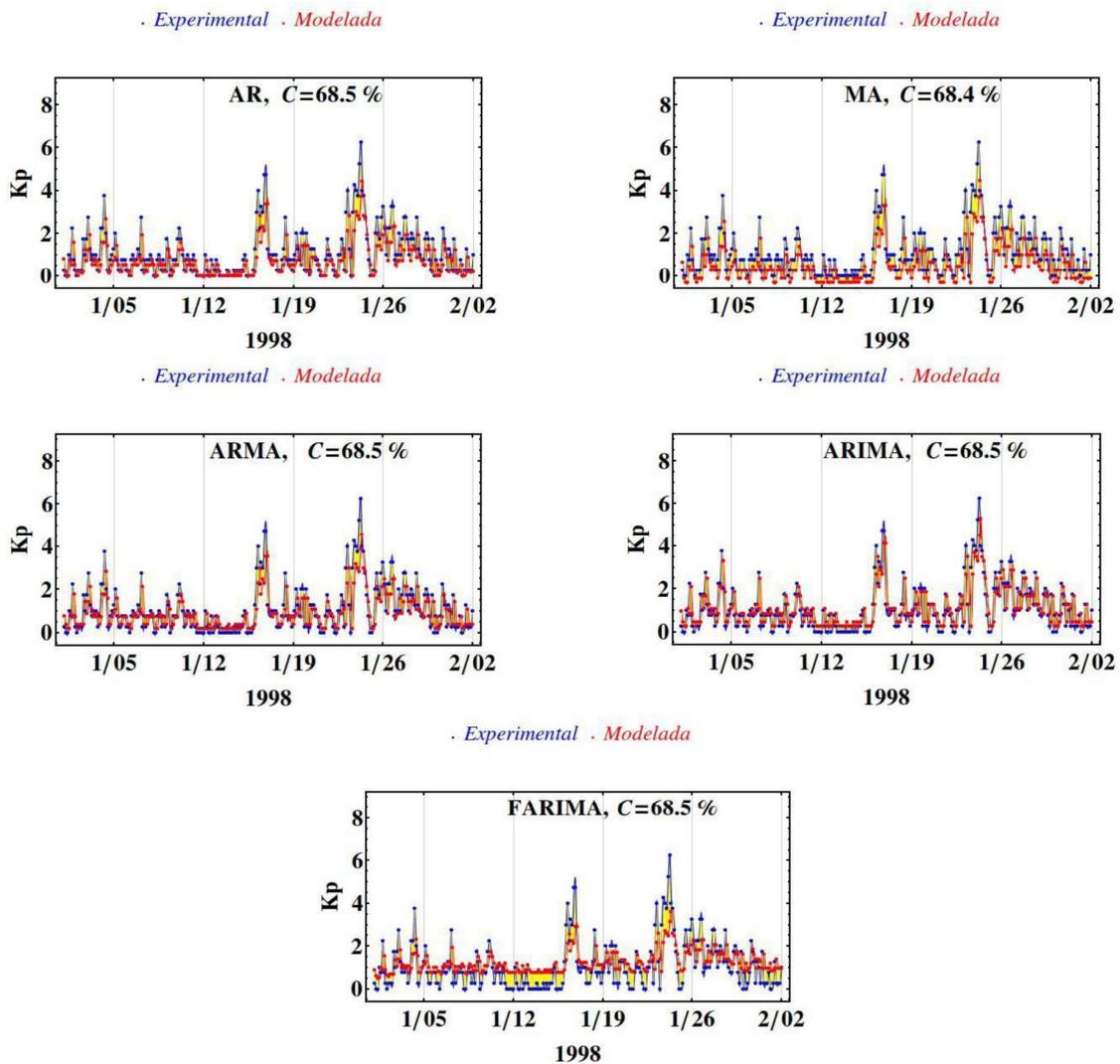

**Fig. 4. En cada panel se representan el Kp real en color azul y el Kp pronosticado en color rojo, ambos obtenidos de la predicción de 3 horas realizado a lo largo de una ventana temporal de 30 días de datos. Se pueden comparar los 5 modelos utilizados en este trabajo, siendo el modelo ARIMA el que muestra mejores resultados para los valores perturbados del índice Kp**

Los cinco modelos propuestos fueron probados y los resultados se representan en cada uno de los paneles de la Figura 5. Sólo se representa la serie histórica del índice Kp desde el 10 de diciembre de 2012, pero se utilizaron todos los datos históricos del Kp desde el año 1932. En color azul tenemos los datos históricos, en color rojo los tres valores pronosticados del Kp y en color verde los valores derivados del cálculo del error medio cuadrático. Este último se le sumó y restó al valor pronosticado en cada hora, así surge la franja semicoloreada la cual representa una franja de validez de los resultados. Los tres puntos de color azul en el interior de la región de predicción son los valores reales del Kp que ya estaban disponibles (en el centro mundial de datos) en el mes de marzo de 2013 cuando el programa fue ejecutado, ellos son Kp=[1+, 0+ e 1+]. Los cálculos de los cinco modelos predicen valores próximos del valor real (puntos de color azul en la Figura 5) en las primeras 3 horas, pero en las restantes 6 horas los modelos MA y ARIMA tienen valores distantes del real.

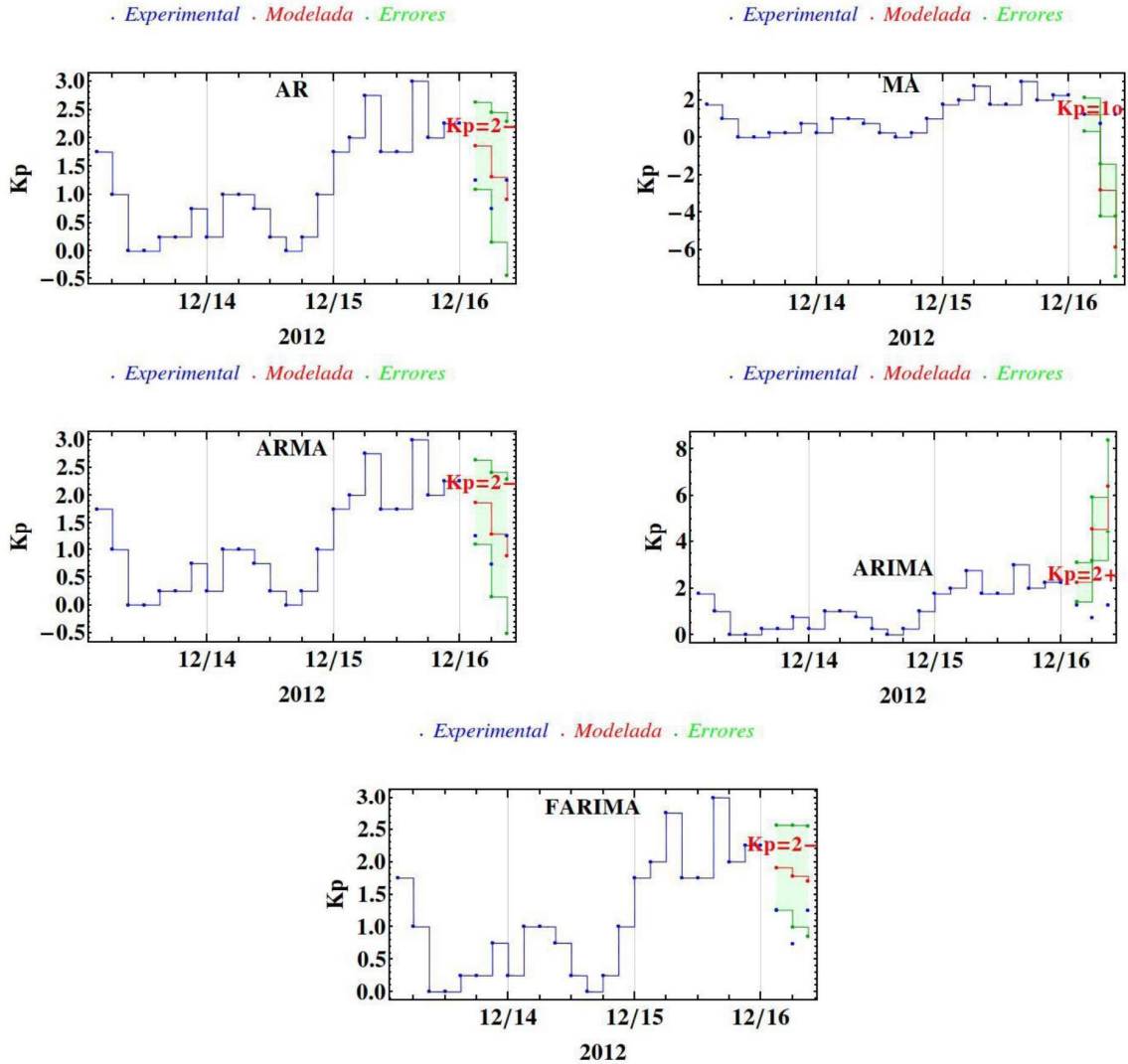

**Fig. 5. Los paneles muestran el pronóstico de 9 horas del índice Kp para los cinco modelos estudiados. El valor del Kp en las próximas 3 horas se muestra en color rojo en el interior de los gráficos**

**Conclusiones**

Durante este trabajo fue propuesto hacer un pronóstico del índice planetario Kp a partir de la serie histórica que comenzó a ser publicada a partir del año 1932. Se confeccionó una base de datos de los índices Kp y del ap y se construyeron histogramas. Posteriormente se procedió a transformar el Kp de *string* a real para poder procesar la serie temporal. Existió el inconveniente de que la escala logarítmica del Kp representada por el valor del ap no era adecuada para hacer el pronóstico porque el tiempo de cálculo era muy lento. Por esta razón se creó una escala lineal. Fueron utilizados cinco modelos autoregresivos, que usan polinomios de interpolación para hacer la predicción de series temporales. Todos los métodos probados son funciones ya implementadas.

Las pruebas realizadas mostraron la utilidad de estos modelos para hacer el pronóstico de 9 horas del índice Kp. El modelo AR se propone como el de menos costo computacional ofreciendo buenos resultados. El modelo ARIMA es eficiente para la predicción del Kp perturbado.

Finalmente este trabajo ofrece una forma rápida y eficiente de hacer una predicción del índice Kp, sin necesidad de usar datos de satélites que muchas veces demoran en ser publicados. Aunque se informa que los resultados del pronóstico son mejores cuando se utilizan datos de satélites. En la literatura se publicó que la correlación lineal entre los valores pronosticados y los valores reales está entorno de un 77 %, valor que es mejor que el 68.5% obtenido en este trabajo. Sin embargo, teniendo en cuenta que se trabajó solamente sobre la serie temporal estocástica del Kp, este valor de correlación puede considerarse satisfactorio.

**Agradecimientos**

**Referencias**

**Boberg, F., Wintoft, P., Lundstedt, H.,** 2000. Real time Kp predictions from solar wind data using neural networks. Physics and Chemistry of the Earth, Part C: Solar, Terrestrial & Planetary Science 25 (4), 275–280.

**Gehred, P., Cliffswallow, W., Schroeder, J., Laboratory, S. E.**, 1995. A Comparison of USAF Ap and Kp Indices to Göttingen Indices. NOAA technical memorandum ERL SEL. U.S. Department of Commerce, National Oceanic and Atmospheric Administration, Environmental Research Laboratories, Space Environment Laboratory.

**McPherron, R. L.,** 1999. Predicting the Ap index from past behavior and solar wind velocity. Physics and Chemistry of the Earth, Part C: Solar, Terrestrial & Planetary Science 24 (1–3), 45–56.

**Shumway, R. H., Stoffer, D. S**., May 2006. Time Series Analysis and Its Applications: With R Examples (Springer Texts in Statistics), 2nd Edition. Springer.

**Takahashi, K., Toth, B. A., Olson, J. V. ,** 2001**.** An automated procedure for near-real- time Kp estimates. Journal of Geophysical Research: Space Physics 106 (A10), 21017–21032.

**Wing, S., Johnson, J. R., Jen, J., Meng, C.-I., Sibeck, D. G., Bechtold, K., Freeman, J., Costello, K., Balikhin, M., Takahashi, K.,** 2005. Kp forecast models. Journal of Geophysical Research: Space Physics 110 (A4), n/a– n/a.

Acerca de los autores:

Dr. Arian Ojeda-González.
Formado en Física Nuclear en el Instituto Superior de Tecnologías y Ciencias Aplicadas, INSTEC, Cuba (2003). Hizo Maestría en Física en la Universidad de la Habana, Cuba (2007). Doctorado en Geofísica Espacial en el Instituto Nacional de Pesquisas Espaciais (2013). Tiene categoría docente de Profesor Instructor dado por la Universidad de la Habana y Aspirante a investigador CITMA/AMA/IGA. Actualmente realiza estudios de Postdoctorado con beca CNPq, en el Instituto Nacional de Pesquisas Espaciais, Brasil.

Dr. Clezio Marcos Denardini
Ingeniero Eléctrico de la Universidad Federal de Santa Maria (UFSM), Brasil (1996). Hizo maestría en Geofísica Espacial en el Instituto Nacional de Pesquisas Espaciais (1999). Doctorado en Geofísica Espacial en el Instituto Nacional de Pesquisas Espaciais (2003). Actualmente es investigador y profesor titular del Instituto Nacional de Pesquisas Espaciais. Presidente da SBGEA (Sociedad Brasileña de Geofísica Espacial y Aeronomía).

MSc. Siomel Savio-Odriozola.
Licenciado en Física en la Universidad de la Habana, Cuba (1999). Hizo Maestría en Física en la Universidad de la Habana, Cuba (2009). Tiene categoría docente de Profesor Instructor por el ITM/Habana/Cuba, es investigador agregado del Departamento de Geofísica Espacial del Instituto de Geofísica y Astronomía, la Habana, Cuba. Actualmente cursa estudios de doctorado en Geofísica Espacial en el Instituto Nacional de Pesquisas Espaciais, Brasil.

Dr. Reinaldo Roberto-Rosa
Formado en Física y Astronomía en la Universidad Federal de Rio de Janeiro, Brasil (1988). Hizo maestría en Astrofísica en el Instituto Nacional de Pesquisas Espaciais (1991). Doctorado en Astrofísica en el Instituto Nacional de Pesquisas Espaciais (1995). Actualmente es investigador y profesor titular del Instituto Nacional de Pesquisas Espaciais, Brasil.

Dr. Odim Mendes-Junior,
Formado en Física en la Universidad Federal de Goiás, Brasil (1982). Hizo maestría en Geofísica Espacial en el Instituto Nacional de Pesquisas Espaciais (1985). Doctorado en Geofísica Espacial en el Instituto Nacional de Pesquisas Espaciais (1992). Actualmente es investigador y profesor titular del Instituto Nacional de Pesquisas Espaciais, Brasil.